\documentclass[sigconf,authorversion,nonacm]{acmart}
\AtBeginDocument{%
  \providecommand\BibTeX{{%
    \normalfont B\kern-0.5em{\scshape i\kern-0.25em b}\kern-0.8em\TeX}}}

\usepackage{multirow}
\usepackage{graphicx}
\usepackage{makecell}
\usepackage{algorithm}
\usepackage{algorithmic}

\begin{document}

\title{Is Meta-Learning the Right Approach for the Cold-Start Problem in Recommender Systems?}

\author{Davide Buffelli}
\authornote{Work done during an internship at Meta AI, Davide is now at MediaTek Research.}
\email{davide.buffelli@phd.unipd.it}
\orcid{0000-0001-5565-1634}
\affiliation{%
	\institution{University of Padova}
	\city{Padova}
	\country{Italy}
}

\author{Ashish Gupta}
\email{vplachouras@meta.com}
\affiliation{%
	\institution{Meta AI}
	\city{London}
	\country{United Kingdom}
}

\author{Agnieszka Strzalka}
\email{vplachouras@meta.com}
\affiliation{%
	\institution{Meta AI}
	\city{London}
	\country{United Kingdom}
}

\author{Vassilis Plachouras}
\email{vplachouras@meta.com}
\affiliation{%
	\institution{Meta AI}
	\city{London}
	\country{United Kingdom}
}

\renewcommand{\shortauthors}{Buffelli et al.}

\begin{abstract}
	Recommender systems have become fundamental building blocks of modern online products and services, and have a substantial impact on user experience. 
	In the past few years, deep learning methods have attracted a lot of research, and are now heavily used in modern real-world recommender systems. 
	Nevertheless, dealing with recommendations in the \textit{cold-start} setting, e.g., when a user has done limited interactions in the system, is a problem that remains far from solved.
	Meta-learning techniques, and in particular optimization-based meta-learning, have recently become the most popular approaches in the academic research literature for tackling the cold-start problem in deep learning models for recommender systems. 
    However, current meta-learning approaches are not practical for real-world recommender systems, which have billions of users and items, and strict latency requirements.
	In this paper we show that it is possible to obtaining similar, or higher, performance on commonly used benchmarks for the cold-start problem \textit{without} using meta-learning techniques. In more detail, we show that, when tuned correctly, standard and widely adopted deep learning models perform just as well as newer meta-learning models. We further show that an extremely simple modular approach using common representation learning techniques, can perform comparably to meta-learning techniques specifically designed for the cold-start setting while being much more easily deployable in real-world applications.
\end{abstract}





\maketitle

\section{Introduction}
Recommender systems are a ubiquitous component of modern online applications and services. To provide some examples, recommender systems are used in entertainment services, to suggest the media items that can be more appealing to a given user, and in e-commerce websites, to recommend items that are likely to be of interest to the user. 
In general, recommender systems are useful when the number of potential items available to the users is extremely large, and hence it is fundamental to tailor the offering to each user in order to ensure a good experience. Other popular examples of this scenario are news outlets and search engines.

In the past few years, deep learning methods have become the standard choice in many domains like computer vision \cite{CHAI2021100134}, natural language processing \cite{khurana2022natural}, and also recommender systems \cite{8529185,10.1145/3285029,zhang2021artificial,10.1145/3535101}. 
Deep learning methods leverage large quantities of data to extract complex relationships between users and items, which have shown to achieve great success in many industrial applications (e.g., \cite{yao2021self,zou2020neural,gu2021self,hashemi2021neural,10.1145/3447548.3467110}). 

While deep learning methods have shown remarkable performance, they are not able to overcome all the challenges that are encountered in applications of recommender systems. One such challenge, which is of great practical importance, and that remains an open research topic, is known as the ``\textit{cold-start problem}''. The cold-start problem generally refers to the setting in which the system is required to make recommendations with limited prior information. This setting can arise in multiple situations, and appears in almost all practical applications (e.g., when the system needs to make recommendations for a  user that has done limited interactions with the system). 
Tackling the cold-start problem is particularly important, as, for example, ensuring a recent user has a positive experience in its  interactions with the system is of paramount importance to ensure the user will continue using the service.

Recently, meta-learning \cite{hospedales2021meta} has become the dominant approach in literature for tackling the cold-start problem in recommender systems \cite{wang2022deep}. The idea behind meta-learning is to \textit{learn the learning process}, i.e., to leverage multiple learning episodes to find the optimal learning algorithm.
This framework is particularly appealing for the few-shot learning setting \cite{song2022comprehensive,bendre2020learning,huang2022survey}, as meta-learning can be used to learn how to learn from a limited number of examples. 

Therefore, in recommender systems, meta-learning can be used to learn how to identify the preferences of a user based on a limited number of interactions, which is similar to the scenario appearing in the cold-start settings.
This parallelism between the few-shot setting and the cold-start setting \cite{lee2019melu}, have led to a wave of research applying meta-learning techniques to recommender systems (e.g., \cite{10.1145/3394486.3403207,10.1145/3404835.3462843,10.1145/3331184.3331268,10.1145/3394486.3403113,Neupane_Zheng_Kong_Yu_2022}).

Meta-learning techniques, like optimization-based meta-learning (which is the most popular in the cold-start problem research), however, present several shortcomings. For example, the training of these methods can require the computation of second-order derivatives, which makes these methods expensive computationally, and difficult to train to convergence \cite{hospedales2021meta,antoniou2018how}. In addition, typical meta-learning methods for the cold-start problem, require to update the parameters of the model (usually through some steps of gradient descent) in order to \textit{adapt} the model to a specific user. These techniques then require the computation of ``adapted'' weights, both at training and at inference time, for every different user. These aspects make current meta-learning methods prohibitive for large scale systems with billions of users, like modern social media and e-commerce platforms, which have strict resource and latency constraints.

Recently, several works in the few-shot learning literature have shown that it is possible to obtain performance comparable, if not superior, to those of complex meta-learning methods, with much simpler baselines that do not use meta-learning \cite{chen2021meta,10.1007/978-3-030-58568-6_16,wang2019simpleshot,chen2018a}. The main trend outlined by these works is that good \textit{representations} are more important on downstream performance than the use of meta-learning.

In this paper, we show that, when properly tuned, widely adopted and relatively ``old'' models like DeepFM \cite{Guo2017DeepFMAF} and DropoutNet \cite{NIPS2017_dbd22ba3} actually perform comparably to recently proposed meta-learning models. Furthermore, we show that a simple modular approach focusing on learning representations that encapsulate all relevant aspects of a given user, can perform comparably, if not better, than meta-learning based models in the cold-start setting.
In more detail, we factorize the representation of a user in five components: (i) user features, (ii)  interactions, (iii) items to be ranked, (iv) social connections, and (v) related users, and we use standard modern techniques such as self-attention \cite{vaswani2017attention}, Deep Sets \cite{NIPS2017_f22e4747}, and graph neural networks \cite{wu2020comprehensive} to learn meaningful representations.
We summarize the contributions of this paper as follows.
\begin{itemize}
    \item We discuss the practical limitations of current meta-learning models specifically designed for the cold-start scenario. 
    \item We show that ``old'' models like DeepFM \cite{Guo2017DeepFMAF} and DropoutNet \cite{NIPS2017_dbd22ba3}, when tuned correctly, perform comparably on popular benchmarks to recent meta-learning methods, which introduce complex and expensive mechanisms that render them impractical for real-world scenarios.
    \item We show that a simple modular framework focused on representation learning, performs comparably, or even better, than recent meta-learning techniques, while being much more easily deployable in real-world settings.
    \item We introduce a new split, focusing on multiple cold-start scenarios, for the recently proposed Kuairec dataset \cite{gao2022kuairec}.
\end{itemize}
Our results show that representation learning is of central importance for tackling the cold-start problem in recommender systems, and that meta-learning techniques, which are not practical for real-world systems, can be matched in performance by much simpler models on current popular cold-start benchmarks.

\section{Preliminaries}
\sloppy In this section we first introduce the main concepts behind recommender systems, the cold-start problem, and meta-learning (with a focus on \textit{optimization-based} meta-learning).
\subsection{Recommender Systems}
Recommender systems can be seen as \textit{ranking} systems, in which a given input query contains information about a user, and the output is a ranked list of items (where a higher ranking means the item is more likely to be of interest to the user) \cite{10.1145/2988450.2988454}.

In general, we have a set of $n$ users $\{u_1, u_2, \dots, u_n\}$, and a set of $m$ items $\{i_1, i_2, \dots, i_m\}$. For each user we might also have a $d_u$-dimensional vector of features $\mathbf{x}^{(u)}_i \in \mathbb{R}^{d_u}, \forall i \in 1,\dots, n$, encoding information like location, number of followers, language, etc.. Similarly, we may have features also for the items $\mathbf{x}^{(i)}_j \in \mathbb{R}^{d_i}, \forall j \in 1,\dots, m$. There is then a set of interactions $\mathcal{I} = \{o_1, o_2, \dots\}$, where every interaction $o$ is a tuple $(u_i, i_j, r)$ where the first element is the user that did the interaction, the second element is the item that the user has interacted with, and the third is the rating that the user has given to the item (this rating may be a number within a certain range, or a binary value for cases of like/dislike or click/no-click).
Typically, there are two ways of encoding the interaction data. 
The first uses an \textit{interaction matrix} $\mathbf{I} \in \mathbb{R}^{n \times m}$, where the element in position $i, j$ contains the rating that user $i$ has given to item $j$.
The second uses a bipartite graph where there is a node for each user and a node for each item, and there is an edge between a user and an item if the user has interacted with the item (the rating is then attached as an edge feature). 
Finally, there may also be information about connections between users, for example in social media two users are connected if they are friends, or if they follow each-other.
Recommender systems are tasked to use the above information to rank the items based on the preference of a given user. 

The most popular approach to recommender systems is that of \textit{collaborative filtering}, which refers to the family of techniques that leverage information from multiple users to inform the predictions for each user. The underlying assumption is that if two users share the same interests, then the interactions of one can be informative also of the preferences of the other.
At a high level, the classical approach \cite{sarwar2000application,koren2009matrix,koren2008factorization,koren2009collaborative} to collaborative filtering is to use matrix factorization techniques on the interaction matrix $\mathbf{I}$ to obtain a representation for each user and item. The dot product of the representation of a user and an item will then provide an estimated score which can be used to rate the items. 
Deep learning approaches instead use neural networks to obtain representations for users and items, and to estimate the ranking of the items that better fits the interests of the user.

\paragraph{The Cold-Start Problem.} 
The cold-start problem refers to the case in which the system is asked to make recommendations with limited data available about the user/item.
For example, common scenarios in which the cold-start appears are:
\begin{itemize}
	\item recommendations are needed for a user that recently started using the service, and there is hence limited information regarding the user's interactions with the system.
	\item the system is asked to recommend to the ``correct'' users an item that has recently been added, hence with limited information of how/which users have interacted with it.
	\item a user has recently come back to the service after a period of inactivity, which may render the past interaction data for this user irrelevant and only limited recent data is available.
	\item a service or system has been recently introduced, and hence there is limited data about its usage.
\end{itemize}

\subsection{Meta-Learning}\label{sec:meta_learn}
Meta-learning refers to techniques that aim at \textit{learning to learn} \cite{vanschoren2018meta}. In more detail, the key idea is to optimize the model parameters \textit{and} the learning process itself by observing multiple learning episodes. 
For this reason, in meta-learning, the input is not a single labelled instance, like in traditional supervised-learning, but it is a \textit{learning episode} $\mathcal{E}_i$. Each learning episode is a tuple composed of a support set and a target set: $\mathcal{E}_i = (\mathcal{S}_{\mathcal{E}_i}, \mathcal{T}_{\mathcal{E}_i})$, where support and target sets are simply sets of labelled examples: 
\begin{align}
	\mathcal{S}_{\mathcal{E}_i} & = \{(x^{(s)}_1, y^{(s)}_1), (x^{(s)}_2, y^{(s)}_2), \dots, (x^{(s)}_{n_s}, y^{(s)}_{n_s})\}, \nonumber \\
	\mathcal{T}_{\mathcal{E}_i} & = \{(x^{(t)}_1, y^{(t)}_1), (x^{(t)}_2, y^{(t)}_2), \dots, (x^{(t)}_{n_t}, y^{(t)}_{n_t})\} \nonumber. 
\end{align}
If we consider the task of image classification as an example, then each $x$ represents an image, and its corresponding $y$ is the label for the image.

Meta-learning techniques usually are composed of a nested learning procedure with an \textit{inner loop} and an \textit{outer loop}. The inner loop is also referred to as the \textit{base-learning} phase, in which a model learns to solve a given task using a learning algorithm and the support set. In the outer loop, also referred as meta-learning, a \textit{meta}-learning algorithm updates the inner learning algorithm, such that the model learned in the inner loop improves its performance on the target set. 

In general, the meta-learning training procedure can be summarised as in Algorithm \ref{alg:metalearning}, which takes as input a model $f$ with parameters $\theta$.
In the pseudocode, \texttt{init} is a method that initializes the parameters $\theta$ (e.g., see \cite{narkhede2022review}). \texttt{ADAPT} represents the inner loop, and is a function that uses the instances in the support set, and a learning algorithm (e.g., simple gradient descent is a common choice) to produce a new version of the parameters $\theta^{\prime}$ adapted to the support set. \texttt{TEST} is a method that computes the loss on the target set of the model $f$ with the adapted parameters $\theta^{\prime}$. Finally, \texttt{UPDATE} is a function that updates the parameters $\theta$ such that the \texttt{ADAPT} method on a given support set leads to good results on the respective target set.

\begin{algorithm}[tb]
	\caption{Meta-learning procedure.}\label{alg:metalearning}
	\begin{algorithmic}
		\STATE {\bfseries Input:} Model $f_{\theta}$; Learning Episodes $\{ \mathcal{E}_1, .., \mathcal{E}_n \}$
		\STATE \texttt{init}$(\theta)$  \; 
		\FOR{$\mathcal{E}_i$  {\bfseries in} $\{ \mathcal{E}_1, .., \mathcal{E}_n \}$}
		         
		\STATE $\theta^{\prime} \leftarrow \text{\texttt{ADAPT}}(f_{\theta}, \mathcal{S}_{\mathcal{E}_i})$ \;
		\STATE $\text{\texttt{o\_loss}} \leftarrow \text{\texttt{TEST}}(f_{\theta^{\prime}}, \mathcal{T}_{\mathcal{E}_i})$ \;
		           
		\STATE $\theta \leftarrow \text{\texttt{UPDATE}}(\theta, \theta^{\prime}, \text{\texttt{o\_loss}})$ \;
		\ENDFOR     
	\end{algorithmic}
\end{algorithm}

When targeting the few-shot setting, the support sets contain a small number of labelled examples, such that the meta-learning procedure leads to a setting of the parameters where a few examples are enough to adapt the model to perform well on a target set of interest. However, this means that also at inference time the model needs to be provided with a labelled support set to be used for adaptation before making predictions on the target set.

It is common to divide meta-learning techniques in three categories (a more detailed taxonomy can be found in \cite{hospedales2021meta}): 
\begin{itemize}
	\item Optimization-based meta-learning \cite{pmlr-v70-finn17a}. These methods rely on second-order (or higher) derivatives to directly optimize through the gradient descent operations in the inner loop.
	\item Model-based meta-learning \cite{qiao2018few}. These are methods in which the output of the inner loop (i.e., the parameters adapted on a support set) is directly provided by a neural network (e.g. a network is used to take as input the support set, and produce as output the parameters to be used on the target set). 
	\item Metric-learning \cite{koch2015siamese}. In these methods the outer loop corresponds to learning a model that outputs representations for the inputs such that the closest training neighbours in the representation space will provide the correct label for the input.
\end{itemize}

\begin{figure*}[h]
\includegraphics[width=\linewidth]{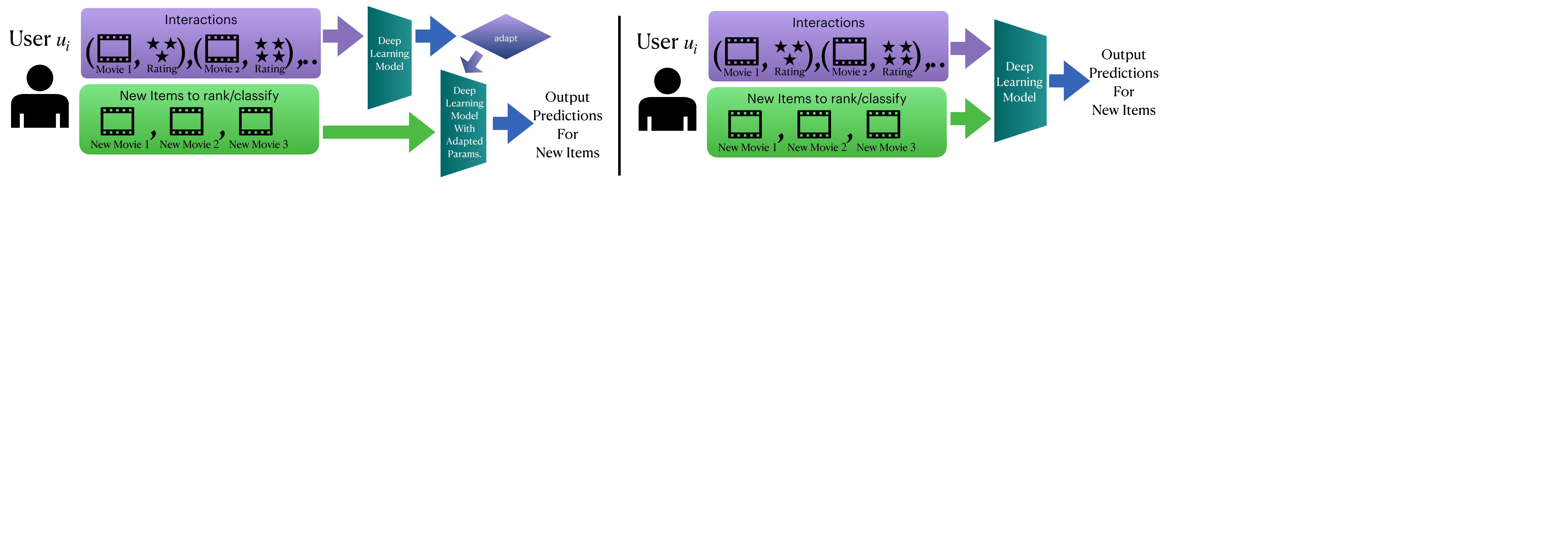}
\caption{Difference in the inference procedure for meta-learning methods (left), and non-meta-learning methods (right). The first have an additional step in which the support set is used to ``adapt'' the parameters to a specific user, which is not required in non-meta-learning methods.}\label{fig:diff_meta_learn}
\end{figure*}

\paragraph{Optimization-Based Meta-Learning}
The first, and most popular, optimization-based meta-learning method is MAML \cite{pmlr-v70-finn17a}, or Model-Agnostic Meta-Learning. The aim of MAML is to find a configuration of the parameters such that a few steps of gradient descent to minimize a loss computed on the support set, can lead the model into a configuration that achieves good performance on the respective target set.

In more detail, referring to the general meta-learning procedure from Algorithm \ref{alg:metalearning}, in MAML the \texttt{ADAPT} function performs some steps (we use $T$ to indicate the number of steps, and $\theta^{\prime}(t)$ to indicate the parameters after $t$ steps) of gradient descent:
\begin{align}
	\theta^{\prime}(0)      & = \theta                                                                                                                            \\
	                        & \cdots  \nonumber                                                                                                                            \\
	\theta^{\prime}_{i} (t) & = \theta^{\prime} (t-1) - \alpha \nabla_{\theta^{\prime} (t-1)} \mathcal{L}(f_{\theta^{\prime} (t-1)}, \mathcal{S}_{\mathcal{E}_i}) \\
	                        & \cdots  \nonumber                                                                                                                             \\
	\theta^{\prime}         & = \theta^{\prime}(T)                                                                                                                
\end{align}
where $\mathcal{L}$ is a loss function computed on the support set (e.g., cross-entropy for classification tasks, or mean squared error for regression tasks), $\alpha$ is the learning rate, and $\nabla_{\theta^{\prime}_{i} (t-1)} \mathcal{L}$ is the gradient of the loss $\mathcal{L}$ with respect to the parameters $\theta^{\prime}_{i} (t-1)$.
The outer loop, is then composed by the following \texttt{UPDATE} function:
\begin{equation}
	\theta = \theta - \beta \nabla_{\theta}\mathcal{L}_{\text{\textit{meta}}} = \theta - \beta \nabla_{\theta} \mathcal{L}(f_{\theta^{\prime}}, \mathcal{T}_{\mathcal{E}_i}) \label{eq:meta_opt}
\end{equation}
where $\beta$ is the learning rate. Notice how in equation \ref{eq:meta_opt}, the loss is computed with the adapted parameters $\theta^{\prime}$, but the gradient is with respect to the initial parameters $\theta$. This leads to a backpropagation through the backpropagation process, which involves second-order derivatives. In equation \ref{eq:meta_opt} we limit ourselves to one step of gradient descent for clarity, but any other optimization algorithm could be used; furthermore, multiple learning episodes can be batched together, in a similar fashion to how multiple examples are batched in traditional supervised learning.

Many variations and improvements of MAML have been proposed over the years, e.g. \cite{nichol2018reptile,raghu2019rapid,grant2018recasting,NEURIPS2019_072b030b}, however, these methods still present some open challenges. In more detail, optimization-based meta-learning can require the computation of second-order derivatives, and the differentiation through many inner loop steps, which makes these methods expensive computationally, and difficult to train to convergence \cite{hospedales2021meta,antoniou2018how}. 
For this reason, the number of inner-loop steps $T$ is usually kept small, which however can limit the generalization capabilities to different episodes \cite{hospedales2021meta}.

\section{The Practicality of Meta-Learning in Recommender Systems}\label{sec:practicality_metalearn}
In recommender systems, meta-learning is generally used to tackle the cold-start problem by creating one learning episode for each user. The support set is then composed of interactions of the user, e.g., items that the users has rated (so that labels are available), while the target set contains the items that the system is asked to rank based on the user preferences (or to predict if the user will like or not).

Regardless of the challenges of optimization-based meta learning mentioned in the previous section, there is one important factor which has not been taken in consideration by the recommender systems literature, and it regards its \textit{practicality}.
Typical real-world applications of recommender systems involve scenarios with hundreds of millions, if not billions, of users and items (e.g., in social media \cite{DLRM19}, e-commerce \cite{7927889}, streaming services \cite{steck2021deep}), and have strict latency requirements, as recommendations need to be provided while the user is engaging with the service. As seen in Section \ref{sec:meta_learn} optimization-based meta learning techniques require backpropagation at inference time. This means that to make a prediction for a user, the model has to first be adapted on a support set for that user, which also implies that separate versions of the parameters (one for every user) needs to be obtained (a scheme of the distinction in the inference procedure between meta-learning and non-meta-learning is shown in Figure \ref{fig:diff_meta_learn}).

A naive application of meta-learning would then require to perform some steps of backpropagation on potentially very large models, every time a prediction for a user is needed, which is infeasible for real-world systems with billions of active users and strict latency requirement. This could be circumvented by instead caching the adapted parameters for each user, but then there is still the problem of saving and retrieving the weights of possibly large models, every time a prediction needs to be made. Furthermore, by having different weights for different users, making use of \textit{batching} to obtain the predictions for multiple users/items with a single inference pass becomes problematic. 
The problem of saving the weights of large models for each user could be tackled by applying the adaptation only on the last few layers, but this would not overcome the difficulties in batching, and the latency introduced by having to load different weights for different users.
Finally, recommender systems in real-world applications are regularly retrained, as trends continuously evolve, and meta-learning is notoriously more difficult and slow to train \cite{hospedales2021meta} which may lead to increased costs.

While meta-learning techniques for the cold-start scenario have nice theoretical motivations, we argue that the current methods are not applicable to real-world scenarios, with billions of users and items, and strict latency and computational requirements. Furthermore, similarly to recent literature in different domains \cite{chen2021meta,10.1007/978-3-030-58568-6_16,wang2019simpleshot,chen2018a}, we show that, even in the cold-start setting, it is possible to obtain performance comparable, or superior, to those of meta-learning methods, with models that \textit{do not} use meta-learning.

\section{A Simple Baseline Modular Framework}
\begin{figure*}[h]
\includegraphics[width=0.9\linewidth]{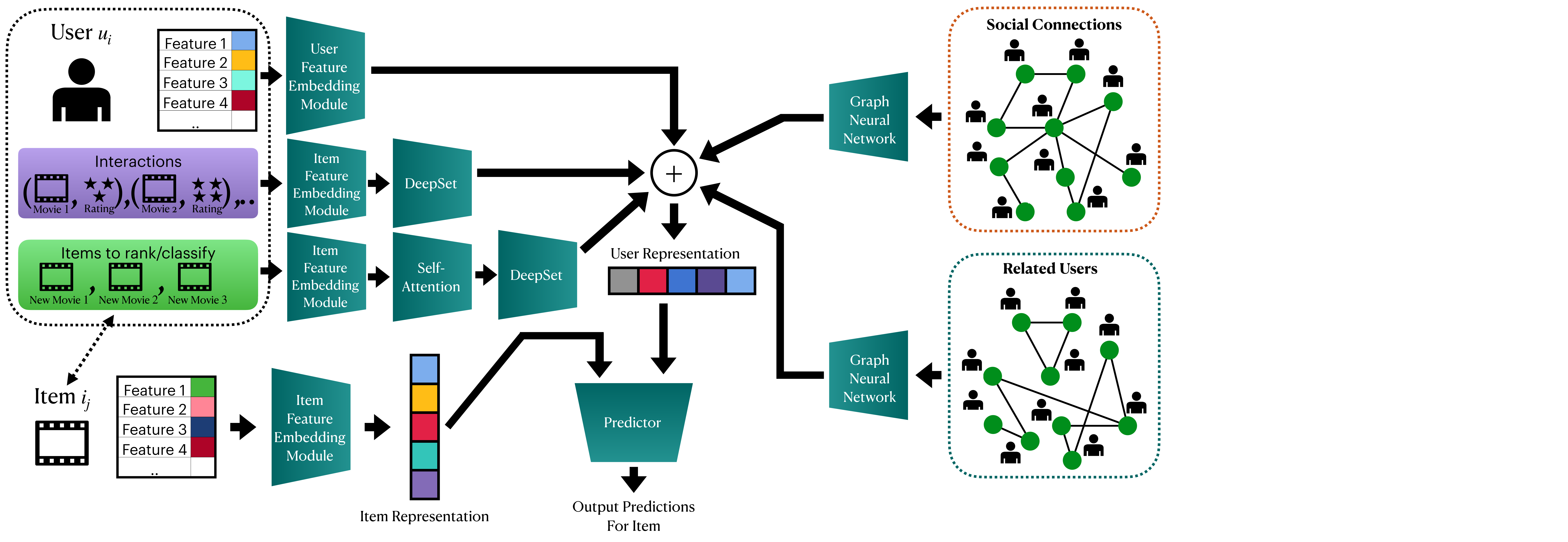}
\caption{High-level illustration of the simple  framework used as baseline for our evaluation. We factorize the representation of a user in five components: user features, interactions, proposed items to be ranked, social connections, and related users (not all components are always available). A predictor then takes as input the representation of a user and an item, and  ranks/classifies the item according to the user's preferences.}\label{fig:our_method}
\end{figure*}
Research on the cold-start problem has shown that auxiliary information coming from features, and social connections, can be leveraged to make better predictions  \cite{BOBADILLA2012225,Shapira2013FacebookSA,tang2013,10.1609/aaai.v33i01.33014189,10.5555/2832747.2832769,ijcai2020p379,sun2011pathsim,7498247,8745499,wang2018ripple}. Furthermore, recent works on few shot learning have shown that good representation learning is more beneficial than sophisticated meta-learning techniques \cite{chen2021meta,10.1007/978-3-030-58568-6_16,wang2019simpleshot,chen2018a}.

In this Chapter we introduce a simple deep learning framework for recommendations that will serve as a baseline model for our evaluation. In more detail, the framework is based on the idea of creating representations encapsulating the information from multiple sources which can help alleviate the cold-start problem, while being simple and modular.  
In fact, we show that by using standard representation learning tools, we can obtain comparable performance to recent and much more complex meta-learning methods designed for the cold-start problem. At a high level, the baseline framework has three main modules: the \textit{user representation module}, the \textit{item representation module}, and the \textit{output prediction module}, and is illustrated in Figure \ref{fig:our_method}.

\subsection{User Representation Module}
The representation of a user plays a particularly important role, as it has to encode information that allows the model to make predictions on the preferences of a user. 
We factorize the representation of a user into multiple components. Each component returns a vector summarizing information from a particular source, and the final representation for the user is obtained by averaging the vectors from the different components. For each dataset, we only use some of the components depending on the data that is available 
(e.g., some datasets may not have all the information needed for some components, e.g., some may not have a social graph). Furthermore, when comparing with other models, we make sure to use only the same information used by these.

The framewoek aim at learning representations for the users that encapsulate information coming from features, interactions (which will be few in the cold-start setting), items to be ranked, social connections, and related users. 
To compare against meta-learning methods, we consider the interactions as the items in the support set, while the items in the target set are the ones to be ranked (exactly as done by meta-learning techniques).

We now describe each of the components in detail.
\paragraph{\textbf{User Features}} We follow the common approach of using embedding tables to obtain a representation for the features of a user \cite{lee2019melu,10.1145/3394486.3403207}. In particular, each feature value is mapped into a $d$-dimensional vector, and finally the vectors for all features are concatenated together.

\paragraph{\textbf{Interactions}} To encode the information from the interactions of a user, we use a DeepSet model \cite{NIPS2017_f22e4747}. Let $\phi$ and $\rho$ be two MLPs, and let $\mathbf{h}_{1}, \mathbf{h}_{2}, \dots, \mathbf{h}_{n_h}$ be the embeddings for the items (obtained as described in Section \ref{sec:item_repr}), the DeepSet model produces the embedding for the interaction data as follows:
\begin{equation}
  \mathbf{E}_h = \rho \left( \sum_i \phi (\mathbf{h}_i) \right)
\end{equation}
While simple, the DeepSet model is proven to be theoretically able to approximate any continuous permutation invariant function over a set \cite{NIPS2017_f22e4747}.

\paragraph{\textbf{Proposed Items to be Ranked}} For processing the items to be ranked\footnote{In modern recommendation systems, the number of items is too large for the model to rank them all for each user, so first there is a mechanism to select a significant subset for each user. To compare against meta-learning methods, this subset is composed of the items that are put in the target set.}, we use a self-attention mechanism \cite{vaswani2017attention}, followed by a DeepSet model. In more detail, let $\mathbf{r}_{1}, \mathbf{r}_{2}, \dots, \mathbf{r}_{n_r}$ be the embeddings for the items, which we stack into a matrix $\mathbf{R}$, and let $\mathbf{W}_q, \mathbf{W}_k, \mathbf{W}_v$ be learnable matrices. The self-attention mechanism proceeds as follows:
\begin{align}
  \mathbf{Q} = \mathbf{R}\mathbf{W}_q,&  \hspace{0.5em} \mathbf{K} = \mathbf{R}\mathbf{W}_k,  \hspace{0.5em} \mathbf{V} = \mathbf{R}\mathbf{W}_v \\
  \mathbf{R}^{\prime} =& \text{ softmax}\left( \frac{\mathbf{Q}\mathbf{K}^{\intercal}}{\sqrt{d^r}} \right)  \mathbf{V}
\end{align}
where $\mathbf{R}^{\prime}$ is the matrix containing the updated representations for the items, $d^r$ is the dimension of the representations, and $\text{softmax}$ is applied row-wise. The intuition behind this strategy is that, with the self-attention mechanism, every item combines information from the other items in a weighted manner, allowing each item to identify the others that are most related to it. This could allow the model to identify ``outliers'', and/or items that are related and should be ranked similarly.
The updated representations for the items are then aggregated into a single vector $\mathbf{E}_r$ using a DeepSet model, as done for the interaction module.

\paragraph{\textbf{Social Connections}} When available, the information from social connections (e.g., if two users are friends on the platform, or if they follow each other) can be used to construct a graph where there is a node for each user and an edge for every connection. We assign a feature vector to each node by taking a concatenation of the features of the user, and of the average of the representations of the items in the  interactions of that user. We then apply a graph convolutional network \cite{kipf2017semisupervised} with multiple layers to learn representations for each user that can capture the relationship between the users and their social connections and preferences.

\paragraph{\textbf{Related Users}} Identifying users that are ``similar'' to each other, but that are not connected by social connections is not an easy task. To keep our baseline simple we rely on heuristics that can be easily applied in practical scenarios. In more detail, we consider two strategies for creating graphs where nodes representing users are connected to each other if the two users are ``similar''.
\begin{description}
    \item{(1)\textit{``Same rating'' graph.}} We connect two users if they have assigned the same rating to at least $k$ items ($k$ is a hyperparameter, but for the considered datasets we set it to 3 without any tuning, in order to show that this simple graph construction already provides useful information, and so that we can obtain information also for cold-start settings). 
    \item{(2)\textit{``Same atrtibute'' graph.}} We connect two users if they share the same value for a specific attribute (multiple of these graphs can be created to consider multiple attribute-induced relationships). We select which attributes to take into account based on the available features. For example, users from the same country.
\end{description}
Each of the created graphs is initialized and processed as done for the \textit{social connections} graph, however, each graph has a separate graph convolutional network, i.e., we do not reuse the same network across graphs, in order to be able to capture different kinds of relations between users.

\subsection{Item Representation Module}\label{sec:item_repr}
As for processing the features of a user, we rely on embedding tables to obtain representations for the items \cite{lee2019melu,10.1145/3394486.3403207}. In particular, each feature value (if a feature is not categorical, then we can always make it categorical by partitioning the space of the values) is mapped into a $d$-dimensional vector, and finally the vectors for each feature are concatenated together.

\subsection{Output Prediction Module}
The Output Prediction Module receives as input the output of the User Representation Module and of the Item Representation Module, and it produces either an estimated score (for ranking), or a probability of like/no like (for classification). We implement this module with a multi-layer perceptron with 2 layers.

\subsection{Availability of Data in the Cold-Start Setting}
As cold-start scenarios entail limited availability of data, we only consider information that is easily accessible even in these cases. 

For the item and user features components, we consider ``fundamental'' features that are always available, even in the cold-start scenario. For example, in an e-commerce service, when a user joins a system, there will always be information about its location and its age. 

Social connection information may not always be available (e.g., in e-commerce websites there usually is none), but in those cases in which it is (e.g., in social networks), it is usually available early after a user starts using the system (at least an initial and incomplete version of it).

For identifying related users we choose heuristics, as described earlier in this Chapter, that can be obtained even in cold-start settings.

Nevertheless, when comparing against other models, we make sure to only use the components of our framework that process information that is used also by those models.

\begin{table*}[h!]
	\caption{Performance of traditional deep learning methods, meta-learning methods, and our simple baseline framework, on four datasets (DBook \cite{10.1145/1060745.1060754}, Movielens \cite{10.1145/2827872}, Yelp \cite{yelp_dataset}, and KuaiRec \cite{gao2022kuairec}). 
 Highlighted are the two best performing models for each dataset and scenario.} 
	\label{main_res}
	\centering
	\resizebox{\linewidth}{!}{%
		\begin{tabular}{cccccc||cccccc}
			\toprule
			\multirow{2}{*}{Scenario}                 & \multirow{2}{*}{Model}                 & \multicolumn{1}{c}{DBook} & \multicolumn{1}{c}{Movielens} & \multicolumn{1}{c}{Yelp} & \multicolumn{1}{c}{KuaiRec} & \multirow{2}{*}{Scenario}                 & \multirow{2}{*}{Model}                 & \multicolumn{1}{c}{DBook} & \multicolumn{1}{c}{Movielens} & \multicolumn{1}{c}{Yelp} & \multicolumn{1}{c}{KuaiRec}\\
			                                          &                                        & nDCG@5                    & nDCG@5                        & nDCG@5                   & ROC AUC                   &  &                                        & nDCG@5                    & nDCG@5                        & nDCG@5                   & ROC AUC    \\
			\midrule
			\multirow{10}{*}{\makecell{Existing items\\for new users\\ (User Cold-start)}} & DeepFM \cite{Guo2017DeepFMAF}                                  & 0.8966 & 0.8414 & 0.8626 & 0.5413 & \multirow{10}{*}{\makecell{New items\\for new users\\ (User-Item Cold-start)}}& DeepFM \cite{Guo2017DeepFMAF}                                 & 0.8879 & 0.8628 & \textbf{0.8403} & 0.5001 \\
			                                          & DropoutNet \cite{NIPS2017_dbd22ba3}    & 0.8792                    & \textbf{0.8999}                        & 0.8424                   &                \textbf{0.7161}            & & DropoutNet \cite{NIPS2017_dbd22ba3}    & 0.8632                    & 0.7794                        & 0.8101                   &                 0.4947            \\   
			\cmidrule{2-6} \cmidrule{8-12}
			                                          & MeLU  \cite{lee2019melu}               & 0.9028                    & 0.8504                        & 0.8719                   &       \textbf{0.6813}                      & & MeLU \cite{lee2019melu}                & 0.8873                    & 0.8588                        & 0.8309                   &  \textbf{0.5761} \\
			                                          & MetaEmb \cite{10.1145/3331184.3331268} & 0.8987                    & 0.8434                        & 0.8447                   &  0.5652                           & & MetaEmb \cite{10.1145/3331184.3331268} & 0.8773                    & 0.8373                        & \textbf{0.8323}                   &             0.5023                \\
			                                          & MWUF \cite{10.1145/3404835.3462843}    & \textbf{0.9288}                    & 0.8957                        & 0.8589                   &     0.5836                        & & MWUF \cite{10.1145/3404835.3462843}    & \textbf{0.9041}                    & 0.8624                        & 0.8274                   &      0.5325                      \\
			                                          & MAMO \cite{10.1145/3394486.3403113}    & 0.8554                    & \textbf{0.9224}                        & \textbf{0.9398}                   &    0.6329                        & & MAMO \cite{10.1145/3394486.3403113}    & 0.8664                    & \textbf{0.9394}                        & 0.7369                   &     \textbf{0.6063}                         \\
                                                      & TSCS \cite{Neupane_Zheng_Kong_Yu_2022} &      0.8792                     &      0.8653                         &     0.8047                     &    0.5519                         &  & TSCS \cite{Neupane_Zheng_Kong_Yu_2022} &   0.8827                        &      0.8211                         &         0.8281                 &            0.5095                 \\
			\cmidrule{2-6} \cmidrule{8-12}
			                                          & Ours (no friends/related users)        & \textbf{0.9272}                    & 0.8517                        & \textbf{0.8867}                   &         0.6340                    & 
                                                      & Ours (no friends/related users)        & \textbf{0.9529}                    & \textbf{0.9902}                        & 0.8254                   &       0.5639                      \\ 
			\midrule
            \multirow{10}{*}{\makecell{New items\\for existing users\\ (Item Cold-start)}} & DeepFM \cite{Guo2017DeepFMAF}                                 & \textbf{0.8814} & 0.\textbf{8578} & 0.8056 & 0.5428 & \multirow{10}{*}{\makecell{Existing items\\for existing users\\ (Non-cold-start)}}& DeepFM \cite{Guo2017DeepFMAF}                                 & \textbf{0.8990} & \textbf{0.8739} & 0.8540 & 0.5221\\
			                                          & DropoutNet \cite{NIPS2017_dbd22ba3}    & 0.8626                    & 0.7407                        & 0.8193                   &                0.5217             & & DropoutNet \cite{NIPS2017_dbd22ba3}    & 0.8866                    & 0.7926                        & \textbf{0.8839}                   &             \textbf{0.8665}               \\
			\cmidrule{2-6} \cmidrule{8-12}
			                                          & MeLU  \cite{lee2019melu}               & 0.8753                    & 0.8189                        & \textbf{0.8477}                   &        0.7507                     &  & MeLU \cite{lee2019melu}                & 0.8900                    & 0.8615                        & 0.8644                   & 0.7261 \\
			                                          & MetaEmb \cite{10.1145/3331184.3331268} & 0.8775                    & 0.8484                        & 0.8311                   &     0.5724                       &  & MetaEmb \cite{10.1145/3331184.3331268} & 0.8856                    & \textbf{0.8806}                        & 0.8660                   &       0.5391                       \\
			                                          & MWUF \cite{10.1145/3404835.3462843}    & 0.8681                    & \textbf{0.8579}                        & 0.8204                   &      \textbf{0.7737}                       &  & MWUF \cite{10.1145/3404835.3462843}    & 0.8965                    & 0.8344                        & 0.8685                   &      \textbf{0.7535}                     \\
			                                          & MAMO \cite{10.1145/3394486.3403113}    & 0.8314                    & 0.7353                        & 0.7784                   &     0.7092                        &  & MAMO \cite{10.1145/3394486.3403113}    & 0.8714                    & 0.8063                        & 0.8652                   &                 0.6994           \\
                                                      & TSCS \cite{Neupane_Zheng_Kong_Yu_2022} &     0.8589                      &     0.8075                          &    0.8113                      &    0.6879                        &   & TSCS \cite{Neupane_Zheng_Kong_Yu_2022} &      0.8862                     &        0.7658                       &          0.8569                &             0.6101                 \\
			\cmidrule{2-6} \cmidrule{8-12}
			                                          & Ours (no friends/related users)        & \textbf{0.9485}                    & 0.7723                        & \textbf{0.8359}                   &          \textbf{0.7632}                   & & Ours (no friends/related users)        & \textbf{0.9114}                    & 0.7196                        & \textbf{0.8899}                   &           0.7229                 \\
            			\bottomrule
		\end{tabular}
	}
\end{table*}

\section{Experiments}
The goal of our experiments is to consider different cold-start scenarios, and compare the performance of recent meta-learning techniques for the cold-start problem with those of traditional deep learning methods for recommender systems, and of our simple baseline framework. The code used for our experiments will be made public upon acceptance.

\begin{table*}
	\caption{Comparison between our simple modular baseline and MetaHIN. Both models use information that goes beyond the interactions of a user.} 
	\label{comparison_metahin}
	\centering
	\resizebox{\linewidth}{!}{%
		\begin{tabular}{cccccc||cccccc}
			\toprule
			\multirow{2}{*}{Scenario}                 & \multirow{2}{*}{Model}                 & \multicolumn{1}{c}{DBook} & \multicolumn{1}{c}{Movielens} & \multicolumn{1}{c}{Yelp} & \multicolumn{1}{c}{KuaiRec} & \multirow{2}{*}{Scenario}                 & \multirow{2}{*}{Model}                 & \multicolumn{1}{c}{DBook} & \multicolumn{1}{c}{Movielens} & \multicolumn{1}{c}{Yelp} & \multicolumn{1}{c}{KuaiRec}\\
			                                          &                                        & nDCG@5                    & nDCG@5                        & nDCG@5                   & ROC AUC                      &                                      &  & nDCG@5                    & nDCG@5                        & nDCG@5                   & ROC AUC        \\
			\midrule
			\multirow{2}{*}{User Cold-start}
			                                          & MetaHIN \cite{10.1145/3394486.3403207} & 0.8957                    & 0.8456                        & \textbf{0.8667}                   &            0.6477                 & \multirow{2}{*}{User-Item Cold-start}
                                                      & MetaHIN \cite{10.1145/3394486.3403207} & 0.8991                    & 0.8640                        & 0.8234                   &       \textbf{0.5518}                      \\ 
			                                          & Ours (rel. users)                  & \textbf{0.9360}                    & \textbf{0.8817}                        & 0.8531                      &           \textbf{0.7538}                  & & Ours (rel. users)                  & \textbf{0.9464}                    & \textbf{0.9530}                        & \textbf{0.8761}                     &  0.5426 \\ 
			\midrule
            \multirow{2}{*}{Item Cold-start}
			                                          & MetaHIN \cite{10.1145/3394486.3403207} & 0.8771                    & 0.8526                        & \textbf{0.8554}                   &            0.6429               & \multirow{2}{*}{Non-cold-start}& MetaHIN \cite{10.1145/3394486.3403207} & \textbf{0.8825}                    & \textbf{0.8584}                        & 0.8535                   &         0.6460                      \\
			                                          & Ours (rel. users)                  & \textbf{0.9188}                    & \textbf{0.8789}                        & 0.8300                       &     \textbf{0.6944}                      &   & Ours (rel. users)                  & 0.8624                    & 0.8088                        & \textbf{0.9052}                       &  \textbf{0.8285}  \\   
			\bottomrule
		\end{tabular}
	}
\end{table*}

\paragraph{Datasets.} For our experiments we choose three of the most popular datasets in the literature of meta-learning methods for the cold-start setting: MovieLens 1M \cite{10.1145/2827872}, DBook \cite{10.1145/1060745.1060754}, and Yelp \cite{yelp_dataset}, and an additional newly released dataset: Kuairec \cite{gao2022kuairec}.
Following previous work \cite{10.1145/3394486.3403207,10.1145/3394486.3403113}, for each dataset, the test data is composed of 4 groups: 
\begin{itemize}
	\item \textbf{User Cold-Start:} this group contains user-item interactions where the users  do not appear in the training data, while the items are present in the training data.
	\item \textbf{Item Cold-Start:} this group contains user-item interactions from users that appear in the training data, but with items that are not present in the training data.
	\item \textbf{User \& Item Cold-Start:} this group contains user-item interactions from users that do not appear in the training data, and that are interacting with items that also do not appear in the training data.
	\item \textbf{Warm (Non Cold-Start):} this group contains user-item interactions from users that appear in the training data, and that are interacting with items that are present in the training data.
\end{itemize}
For MovieLens, DBook, and Yelp, we use the same group division as in prior work \cite{10.1145/3394486.3403207,10.1145/3394486.3403113}, while for Kuairec we manually process the data  to obtain the above groups (additional details are provided in Appendix \ref{appendix:Kuairec}). The code used to process Kuairec, and the pre-processed data, will be made available upon acceptance.

The meta-learning techniques that have been proposed for the cold-start problem in recommender system always require a support set (containing the, possibly small-sized, set of items that the user has interacted with) and a target set (containing the items which the system is required to rank for the user) as input. This implies that we are never considering a ``drastic'' cold-start scenario, e.g., the very first time the user uses the system, but a cold-start case with a very small amount of interactions including the user/item. 

\paragraph{Considered Models.} 
For the non-meta-learning models, we consider two widely adopted deep learning models: DeepFM \cite{Guo2017DeepFMAF}, and DropoutNet \cite{NIPS2017_dbd22ba3}, in addition to our simple baseline framework.
The literature on meta-learning techniques applied to the cold-start problem in recommendations is large and growing, so while it would not be possible to compare against all methods, we choose six representative models among the most popular: MeLU \cite{lee2019melu}, MetaHIN \cite{10.1145/3394486.3403207}, MWUF \cite{10.1145/3404835.3462843}, MetaEmb \cite{10.1145/3331184.3331268}, MAMO \cite{10.1145/3394486.3403113}, and TSCS \cite{Neupane_Zheng_Kong_Yu_2022}.

\paragraph{Implementation \& Hyperparameters.} 
For the considered models, when available, we directly take the implementation of the model released by the authors, or the implementation available in the open source repository Recbole \cite{recbole,recbole[1.1.1],recbole[2.0]}. Some of the models where originally implemented for the task of click-through rate prediction, so we modify them for the task of rating prediction simply by replacing the loss function and the activation function at the output layer of the model (i.e., we change from sigmoid to ReLU).  

For MovieLens, DBook, and Yelp, we follow the same procedure outlined in prior work and train the models to minimize the mean squared error between the prediction score and the true scores that a user has assigned to an item, plus any additional loss or regularization term defined by the authors of each model. 
Differently, for Kuairec the the task is to classify if the user will like or not like a video, and so the models are trained to minimize a binary cross-entropy loss (plus any additional loss or regularization term defined by the authors of each model).
The only exception is DropoutNet, which is trained with a loss function inspired by denoising autoencoders, as proposed by the authors of the model \cite{NIPS2017_dbd22ba3}. 
We train each model with early stopping on the validation set (i.e., as final weights, we take the ones at the epoch with the lowest loss on the validation set).

When available, we use the optimal hyperparameters provided by the authors of each model. For all datasets for which there are no optimal hyperparameters released by the authors, we perform a hyperparameter tuning procedure based on a grid search (more details can be found in Appendix \ref{appendix:baselines_hyperparam}).
Similarly, for identifying the best hyperparamters for our baseline, we perform a simple grid search (the considered values are reported in Appendix \ref{appendix:baselines_hyperparam}) and select the configuration with the lowest loss on the validation set.

\paragraph{Fair Comparison of Meta-Learning and Non-Meta-Learning Methods.} Meta-learning methods receive as input a support and target set, where the support set contains labelled examples that are used to ``adapt'' the parameters of the model (see Section \ref{sec:meta_learn}). This procedure also happens at test-time, meaning that meta-learning models have access to labels that non-meta-learning methods do not have access to, which makes a comparison between the two not fair. Previous works have then adopted the strategy of \textit{fine-tuning} a trained non-meta-learning model using the contents of the support sets in the test data \cite{10.1145/3394486.3403207}. As fine-tuning adds additional hyperparameters, we instead add the labels for the items in the support sets as additional input features for non-meta-learning methods\footnote{We modify DeepFM and DropoutNet so that they receive as input a support set and a target set for a user.}. This allows us to ensure both meta-learning and non-meta-learning models observe the same labels (which is necessary for a fair comparison), without requiring additional hyperparameters or optimization procedures for non-meta-learning models.

\paragraph{Comparison with our Baseline Framework.} Our baseline framework has a modular approach considering different possible components. However, while all models rely on information coming from the features of users and items, and on interaction data, not all the considered models access information from social connections, or related users. For this reason, when comparing our baseline against DeepFM \cite{Guo2017DeepFMAF}, DropoutNet \cite{NIPS2017_dbd22ba3}, MeLU \cite{lee2019melu}, MWUF \cite{10.1145/3404835.3462843}, MetaEmb \cite{10.1145/3331184.3331268}, MAMO \cite{10.1145/3394486.3403113}, and TSCS \cite{Neupane_Zheng_Kong_Yu_2022}, we do not include the modules processing social connections and related users.
MetaHIN \cite{10.1145/3394486.3403207} instead uses information from related users, and hence in this case we also consider the module processing information from related users in our baseline.

\paragraph{Evaluation Metrics.} For MovieLens, DBook, and Yelp, where the model is asked to rank the items based on the preferences of a user, we measure the performance of the models using the normalized Discounted Cumulative Gain (nDCG), as is standard in literature. The nDCG@k intuitively measures the quality of the top $k$ recommendations made by the model, or, in other words, it measures if the $k$ items to which the model assigns the highest score, are close to being the top $k$ items according to the real scores assigned by the user (and further takes the ordering into account). 
For Kuairec, where the model has to predict if a user will like a given video, we use the ROC AUC score, as is standard in binary classification tasks.

\subsection{Results}
In Table \ref{main_res} we report a comparison of the performance of traditional deep learning methods for recommender systems (DeepFM \cite{Guo2017DeepFMAF}, DropoutNet \cite{NIPS2017_dbd22ba3}), meta-learning methods for the cold-start problem (MeLU \cite{lee2019melu}, MWUF \cite{10.1145/3404835.3462843}, MetaEmb \cite{10.1145/3331184.3331268}, MAMO \cite{10.1145/3394486.3403113}, TSCS \cite{Neupane_Zheng_Kong_Yu_2022}), and simple baseline framework. 
We first notice that DeepFM and DropoutNet, while not being specifically designed for the cold-start scenario and being much more simple and dated than meta-learning methods, tend to perform comparably to the latter. This highlights that with proper hyperparameter tuning even more dated and simple methods can achieve the same performance of complicated meta-learning methods, which are not practical for real-world scenarios (as discussed in Section \ref{sec:practicality_metalearn}).

Furthermore, we observe that in more than half of all scenarios and datasets, our simple modular approach is either the best, or second-best, performing model. Furthermore, in $30\%$ of the settings the simple baseline framework is actually the best performing model. 
These results shows that by designing a model that is capable of creating  representations using multiple sources of information, it possible to tackle the cold-start problem without requiring the use of meta-learning methods, which are significantly more challenging to deploy in real-world scenarios.

We also notice that, in the ``user-item cold-start'' scenario, which is the most challenging scenario, involving users and items that were not observed during training, in 3 out of 4 scenarios the highest performing model is not based on meta-learning. This is a very strong indication of the fact that representation learning plays a crucial role in the cold-start problem, and that meta-learning is not necessarily the right solution.

In Table \ref{comparison_metahin} we compare our modular baseline with MetaHIN \cite{10.1145/3394486.3403207}. In this case, both models access information that goes beyond just the features of users and the items they interacted with. We observe that in most cases the baseline leads to higher results. This again highlights that the extra complexity of meta-learning is not providing any benefit, and that a simple modular approach using standard representation learning techniques leads to similar, if not better, results.

\begin{table*}
	\caption{Study of the impact of information extracted from social connections and related users on the performance of our proposed model  on four datasets (DBook \cite{10.1145/1060745.1060754}, Movielens \cite{10.1145/2827872}, Yelp \cite{yelp_dataset}, and KuaiRec \cite{gao2022kuairec}). ``N/A'' indicates that the dataset does not contain the information needed for a specific configuration of our model. Highlighted are the two best performing configurations for each dataset and scenario.} 
	\label{our_ablation}
	\centering
	\resizebox{\linewidth}{!}{%
		\begin{tabular}{cccccc||cccccc}
			\toprule
			\multirow{2}{*}{Scenario}                 & \multirow{2}{*}{Model}                 & \multicolumn{1}{c}{DBook} & \multicolumn{1}{c}{Movielens} & \multicolumn{1}{c}{Yelp} & \multicolumn{1}{c}{KuaiRec} & \multirow{2}{*}{Scenario}                 & \multirow{2}{*}{Model}                 & \multicolumn{1}{c}{DBook} & \multicolumn{1}{c}{Movielens} & \multicolumn{1}{c}{Yelp} & \multicolumn{1}{c}{KuaiRec}\\
			                                          &                                        & nDCG@5                    & nDCG@5                        & nDCG@5                   & ROC AUC                      &                                      &  & nDCG@5                    & nDCG@5                        & nDCG@5                   & ROC AUC        \\
			\midrule
			\multirow{8}{*}{\makecell{Existing items\\for new users\\ (User Cold-start)}}
			                                          & Ours (no friends/related users)        & 0.9272                    & 0.8517                        & \textbf{0.8867}                   &         0.6340                    & \multirow{8}{*}{\makecell{New items\\for new users\\ (User-Item Cold-start)}}
                                                      & Ours (no friends/related users)        & \textbf{0.9529}                    & \textbf{0.9902}                        & 0.8254                   &       \textbf{0.5639}                      \\ 
			                                          & Ours (friends)                         & 0.8963                    & N/A                           & 0.8830                   &               \textbf{0.7664}              & & Ours (friends)                         & 0.8404                    & N/A                           & \textbf{0.8696}                   & \textbf{0.6528} \\    
			                                          & Ours (rel. users: same attr.)          & 0.8859                    & \textbf{0.9063}                        & N/A                      &          0.6729                   & & Ours (rel. users: same attr.).         & 0.8174                    & 0.9339                        & N/A                      & 0.4988 \\   
			                                          & Ours (rel. users: same ratings)        & 0.8760                    & 0.8598                        & 0.8531                   &        0.6739                    &  & Ours (rel. users: same ratings)        & 0.8609                    & 0.8646                        & \textbf{0.8761}                   & 0.5320 \\   
			                                          & Ours (friends + ratings)               & 0.9286                    & N/A                           & \textbf{0.8867}                   &    0.6669                       &  & Ours (friends + ratings)               & 0.9418                    & N/A                           & 0.8465                 & 0.5225  \\   
			                                          & Ours (friends + attr.)                 & \textbf{0.9295}                    & N/A                           & N/A                      &      0.7300                      &  & Ours (friends + attr.)                 & 0.9193                    & N/A                           & N/A                      &  0.5296 \\  
			                                          & Ours (rating + attr.)                  & \textbf{0.9360}                    & \textbf{0.8817}                        & N/A                      &           \textbf{0.7538}                  & & Ours (rating + attr.)                  & \textbf{0.9464}                    & \textbf{0.9530}                        & N/A                      &  0.5426 \\ 
			                                          & Ours (friends + rating + attr.)        & 0.9229                    & N/A                           & N/A                      &     0.6553                      &  & Ours (friends + rating + attr.)        & 0.9072                    & N/A                           & N/A                      & 0.5554   \\   
			\midrule
            \multirow{8}{*}{\makecell{New items\\for existing users\\ (Item Cold-start)}}
			                                          & Ours (no friends/related users)        & \textbf{0.9485}                    & 0.7723                        & \textbf{0.8359}                   &          \textbf{0.7632}                   & \multirow{8}{*}{\makecell{Existing items\\for existing users\\ (Non-cold-start)}}& Ours (no friends/related users)        & \textbf{0.9114}                    & 0.7196                        & \textbf{0.8899}                   &           0.7229                 \\ 
			                                          & Ours (friends)                         & 0.9122                    & N/A                           & 0.8049                   &     \textbf{0.7612}                      &   & Ours (friends)                         & 0.8443                    & N/A                           & 0.8879                   &  \textbf{0.8092}  \\    
			                                          & Ours (rel. users: same attr.)          & 0.9317                    & 0.8033                        & N/A                      &                0.6264             & & Ours (rel. users: same attr.)          & 0.8559                    & \textbf{0.8065}                        & N/A                      & 0.5047\\   
			                                          & Ours (rel. users: same ratings)        & 0.9230                    & \textbf{0.8650}                        & \textbf{0.8300}                   &   0.6251                        &  & Ours (rel. users: same ratings)        & 0.8153                    & 0.7556                        & \textbf{0.9052}                   &  0.5567 \\   
			                                          & Ours (friends + ratings)               & 0.9121                    & N/A                           & 0.8272                  &        0.5372                    &  & Ours (friends + ratings)               & \textbf{0.9055}                    & N/A                           & 0.8872                   & 0.5592 \\   
			                                          & Ours (friends + attr.)                 & 0.9208                    & N/A                           & N/A                      &     0.7094                       &   & Ours (friends + attr.)                 & 0.8891                    & N/A                           & N/A                      & 0.6678 \\   
			                                          & Ours (rating + attr.)                  & 0.9188                    & \textbf{0.8789}                        & N/A                      &     0.6944                      &   & Ours (rating + attr.)                  & 0.8624                    & \textbf{0.8088}                        & N/A                      &  \textbf{0.8285}  \\   
			                                          & Ours (friends + rating + attr.)        & \textbf{0.9414}                    & N/A                           & N/A                      &      0.7560                      &       & Ours (friends + rating + attr.)        & 0.9031                    & N/A                           & N/A                      & 0.7275 \\  
			\bottomrule
		\end{tabular}
	}
\end{table*}

\paragraph{Ablation Study - Impact of Social Connections and Related Users.} We perform a study to understand the impact of the information extracted from social connections, and graphs obtained by identifying related users. Results are shown in Table \ref{our_ablation} where we report the performance of different versions of our simple modular framework. In particular, we consider a version using no information from social connections and related users (the same used for Table \ref{main_res}), one using only social connections, one using only related graphs (as used in Table \ref{comparison_metahin}), and then all possible combinations of social connections and related graphs\footnote{For Movielens there is no social connections data available, so we cannot create a social graph. For Yelp the features are not relevant for identifying ``related'' users, so we do cannot consider graphs of this kind.}. In all cases we always use information from user features,  interactions and items to be ranked, we just avoid considering information from ``graph'' data. 
We notice that in the majority of cases, using information from social connections and related users leads to the highest performance, however the configuration that does not use these sources of information still performs surprisingly well. This shows that properly modelling the information from the features and interactions of a user already provides significant amount of information regarding the preferences of a user in these commonly used datasets.
Furthermore, we notice that the best performing configuration changes across datasets and scenarios. This indicates that it is needed to identify the best configuration for each dataset separately, and that different cold-start scenarios have different peculiarities which should be treated differently. 

\paragraph{Ablation Study - Impact of User Features, and Historical/Recent Interactions}
In Table \ref{tab:no_feat_int} we study how the performance of baseline is impacted when we construct the representation for a user without considering the user's features, the user's  interactions, or the aggregation of the features of items to be ranked. For space limitations, we present the results only on the Yelp dataset as representative, as it is the largest dataset considered. We notice that removing the user features causes a significant drop in performance in all cold-start scenarios, but actually improves performance in the non-cold-start setting.
When removing the contribution of the interactions, we notice that performance on the \textit{item cold-start} and the \textit{user-item cold-start} settings drops significantly, but performance on the \textit{user cold-start} and \textit{non-cold-start} settings increases. 
These results confirm that there is no ``perfect'' configuration that can obtain the highest performance on all cold-start scenarios, but using all components seems to lead to the more ``equilibrated'' configuration that can perform well in all scenarios.
\begin{table}
	\caption{Performance of our model on the Yelp dataset when removing the user feature component (indicated with ``No user feat.''), when removing the interactions component (indicated with ``No inter.''), and when removing the component related to the aggregation of items to be ranked (indicated with ``No ranking agg.'').} 
	\label{tab:no_feat_int}
	\centering
	\resizebox{\linewidth}{!}{%
		\begin{tabular}{ccc||ccc}
			\toprule
			\multirow{2}{*}{Scenario}                 & \multirow{2}{*}{Model}                 & \multicolumn{1}{c}{Yelp} & \multirow{2}{*}{Scenario}                 & \multirow{2}{*}{Model}                 & \multicolumn{1}{c}{Yelp} \\
			                                          &                                        & nDCG@5         &                                      &  & nDCG@5    \\
			\midrule
			\multirow{4}{*}{\makecell{Existing items\\for new users\\ (User Cold-start)}}
			                                          & No user feat.        & 0.8159                 & \multirow{4}{*}{\makecell{New items\\for new users\\ (User-Item Cold-start)}}
                                                      & No user feat.        & 0.7791         \\ 
			                                          & No inter.                        & 0.9048                  & & No inter.                         & 0.8022  \\   
                                                      & No ranking agg.                       & 0.9059                   & & No ranking agg.                        & 0.8150   \\   
                                                      & Full                        & 0.8867                  & & Full                        & 0.8465   \\  
			\midrule
            \multirow{4}{*}{\makecell{New items\\for existing users\\ (Item Cold-start)}}
			                                          & No user feat.        & 0.7910     & \multirow{4}{*}{\makecell{Existing items\\for existing users\\ (Non-cold-start)}}& No user feat.        & 0.9609                     \\ 
			                                          & No inter.                           & 0.7758        &   & No inter.                          & 0.9218         \\    
                                                      & No ranking agg.                        & 0.7529                   & & No ranking agg.                       & 0.9218   \\   
                                                      & Full                        & 0.8272                   & & Full                        & 0.8872   \\  
			\bottomrule
		\end{tabular}
	}
\end{table}

\section{Related Work}
\paragraph{Recommender Systems.}
Research on recommender systems started in the 90s \cite{belkin1992information,goldberg1992using} with the introduction of the concept of \textit{collaborative filtering}, which is still at the base of modern recommender systems. Collaborative filtering refers to techniques that leverage information gathered from multiple users in order better target the preferences of a given user. Before the ``deep learning revolution'', the most popular approaches for recommender systems were based on matrix factorization methods (e.g., \cite{sarwar2000application,koren2009matrix,koren2008factorization,koren2009collaborative}). 
These methods have also been enhanced by incorporating machine learning models such as SVMs (e.g., \cite{rendle2010factorization}) and neural networks (e.g., \cite{Guo2017DeepFMAF}). From 2016, the successes of deep learning in domains like computer vision and natural language processing, have led to the first recommender systems based on neural networks \cite{shan2016deep,covington2016deep}. As it would not be possible to cover the large literature on deep learning models for recommender systems, we refer the interested reader to recent surveys \cite{zhang2019deep,da2020recommendation,khan2021deep,wu2021survey}.
Furthermore, a breakdown of the history of recommender systems can be found in Jannach et al.\ \cite{jannach2021recommender}, and Dong et al.\ \cite{https://doi.org/10.48550/arxiv.2209.01860}.

\paragraph{The Cold-Start Problem.} 
Given the importance and the ubiquity of the cold-start problem, a large amount of research has been done on this topic. At a high level, we can identify two main approaches \cite{8229786}: one directly queries the user for information, while the other implicitly relies on available additional information. 
The first strategy aims at directly asking the users to rate some items, and using these ratings to personalize the recommendations using active learning (e.g., \cite{10.1145/2505515.2505690,10.1007/978-3-319-10491-1_12}) or interview-based approached (e.g., \cite{10.1145/2433396.2433451}).
The second strategy relies on meta-learning (which we discuss in the next paragraph) or on leveraging additional information (e.g., \cite{BOBADILLA2012225,Shapira2013FacebookSA,tang2013}), which can include features (e.g., \cite{10.1609/aaai.v33i01.33014189,10.5555/2832747.2832769,ijcai2020p379}), and information from social connections (e.g., \cite{sun2011pathsim,7498247,8745499}), or other types of graphs (e.g., \cite{wang2018ripple}).
Several surveys on the cold-start problem are available for a more thorough coverage of the topic \cite{8229786,sethi2021cold,abdullah2021eliciting}.

\paragraph{Meta-Learning applied to the Cold-Start Problem.}
In addition to the models we considered in our experiments (MeLU \cite{lee2019melu}, MetaHIN \cite{10.1145/3394486.3403207}, MWUF \cite{10.1145/3404835.3462843}, MetaEmb \cite{10.1145/3331184.3331268}, MAMO \cite{10.1145/3394486.3403113}, TSCS \cite{Neupane_Zheng_Kong_Yu_2022}), there are additional works using meta-learning for the cold-start problem.
Vartak et al.\ \cite{NIPS2017_51e6d6e6} focus on the item cold-start scenario, and use the user's history as support set from which they obtain adapted parameters that are used to make predictions for new items (which compose the target set).
Du et al.\ \cite{10.1145/3292500.3330726} focus on sequential recommendations, and propose a meta-learner composed of an initializer that initializes the parameters of the prediction network, and a controller, which updates the parameters based on new data and decides when to stop the adaptation process.
Wei et al.\ \cite{9338389} use meta-learning to quickly adapt the representation of a new user based on a subgraph of its connected users. 
Xie et al.\ \cite{https://doi.org/10.48550/arxiv.2105.03686} separately model long-term and short term history for a user and use meta-learning to quickly adapt the user representations.
Bharadhwaj \cite{8852100} uses meta-learning to find an initial setting of the parameters that can be shared and quickly adapted to new users.
Finally, we mention that meta-learning has also been used to perform model-selection for recommender systems \cite{CUNHA2018128,10.1007/978-3-319-46227-1_25,10.1145/2365952.2366002}.
A complete survey on the use of meta-learning in recommender systems is provided by Wang et al.\ \cite{wang2022deep}.

\paragraph{Meta-Learning.}
Meta-learning has quite a long history \cite{schmidhuber1987evolutionary,bengio2013optimization,Thrun1998}, but in the past few years some seminal works have cemented its popularity in the few-shot learning setting \cite{ravi2017optimization,NIPS2017_cb8da676,NIPS2016_90e13578}.
MAML \cite{pmlr-v70-finn17a} in particular is the most popular approach, with applications in several domains, like computer vision \cite{Perez-Rua_2020_CVPR}, robotics \cite{10.5555/3327345.3327436}, and language modelling \cite{huang-etal-2018-natural}.
A recent complete review of the meta-learning field is provided by Hospedales et al.\ \cite{hospedales2021meta}.

\section{Conclusion}
Recommender systems are core components of modern web-based applications, and a lot of academic and industrial research has gone into this topic. An important problem that arises in almost all practical scenarios, is that of \textit{cold-start}, i.e., when the system is asked to make recommendations based on limited interaction data about a user and/or item.
Meta-learning has become the most popular method for tackling this issue in the academic research community, however, current methods are ill-suited for real-world large-scale applications.
In this paper, we show that it is possible to obtain performance comparable, and in many cases superior, to those of meta-learning methods, without using meta-learning.

\begin{acks}
    Work done while Davide Buffelli was an intern at Meta AI.
	Davide Buffelli was partially supported by the Italian Ministry of Education, University and Research (MIUR), under PRIN Project n. 20174LF3T8 ``AHeAD'' and the initiative ``Departments of Excellence'' (Law 232/2016), and by University of Padova under project ``SID 2020: RATED-X''. 
\end{acks}

\bibliographystyle{ACM-Reference-Format}
\bibliography{refs}

\appendix

\section{Kuairec Processing}\label{appendix:Kuairec}
The Kuairec dataset \cite{gao2022kuairec} is obtained using data from the Kuaishou mobile app (which is a short-video sharing social network).
The dataset is composed of user-item interaction data, where the items are videos from the platform. In more detail, the dataset has data from 7176 users and 10728 items, with additional information regarding social connections (i.e., friendships between users), a list of tags for each item, and user features\footnote{There is also additional \textit{daily} information about the items, which however we do not use.}.

\paragraph{User-item interaction data.} The user-item interaction data is peculiar, as it contains a subset of 1411 users and 3327 items which is \textit{fully-observed}, i.e., in this subset, all users have interacted with all items. 
For every interaction between a user and an item, there is a set of features which includes how long the user has watched the video. Following the recommendation from the authors of the dataset, we assign a positive label (or, in other words, a ``like''), if the user has spent an amount of time on the video which is greater or equal than double the length of the video.

\paragraph{Social connections.} Users can follow each other on the platform, and this kind of information can be used to produce a graph in which nodes represent users, and edges represent connections between users.

\paragraph{Item Categories.} For each item there is a list of categories, or ``tags'', assigned to it. In total there are 31 different tags. We encode the tags for each item into a one-hot vector and use it as feature vector for the item.

\paragraph{User Features.} Multiple features are available for each user. We use the encrypted one hot features provided in the dataset, which combined lead to a feature vector of $1555$ elements per each user.

\subsection{Partitioning the Dataset}
We obtain the split for our experiments (which we will release in addition to our code upon acceptance) with the following procedure.
\begin{enumerate}
    \item We randomly select $10\%$ of the items in the \textit{fully-observed} subset, and use all the interactions between the users in the \textit{fully-observed} subset and these items to compose the ``Item Cold-Start'' setting. We remove all interactions with these items for all the next steps of this procedure.
    \item We randomly select $10\%$ of the users in the \textit{fully-observed} subset. We use all the interactions from these users to compose the ``User Cold-Start'' setting. We remove all interactions from these users for all the next steps in this procedure.
    \item We randomly select $10\%$ of the remaining interactions in the \textit{fully-observed} subset to obtain the ``Non Cold-Start'' setting. We then remove the selected interactions, and we keep all the remaining ones as training set.
    \item We use the users and items interactions that do not appear in the \textit{fully-observed} subset of the dataset for the ``User-Item Cold-Start'' setting.
\end{enumerate}

\begin{table}
	\caption{Values used for the hyperparamter tuning of each considered meta-learning model.} 
	\label{tab:baseline_hyp}
	\centering
	\resizebox{\columnwidth}{!}{%
	\begin{tabular}{lcc}
		\toprule
		Model                                  & Hyperparameter & Values \\
		\midrule
		\multirow{9}{*}{MeLU \cite{lee2019melu}} &  Hidden Layers Size & $32, 64, 128$ \\   
                                                                &  Embedding Dimension & $32, 64, 128$ \\   
                                                                &  User Embedding Dimension & $32, 64, 128$ \\
                                                                &  Item Embedding Dimension & $32, 64, 128$ \\
                                                                &  Inner Loop Steps (Training) & $1, 2, 4$ \\
                                                                &  Inner Loop Learning Rate & $0.05, 0.01, 0.1, 0.001$ \\ 
                                                                &  Outer Loop Learning Rate & $0.01, 0.005, 0.001, 0.0005$ \\
                                                                & Epochs & $30, 40, 50, 60$ \\
                                                                & Batch Size & $10, 20$ \\
        \midrule
		\multirow{11}{*}{MetaHIN \cite{10.1145/3394486.3403207}} &  Hidden Layers Size & $32, 64, 128$ \\   
                                                                &  Embedding Dimension & $32, 64, 128$ \\   
                                                                &  User Embedding Dimension & $32, 64, 128$ \\
                                                                &  Item Embedding Dimension & $32, 64, 128$ \\
                                                                &  Inner Loop Steps (Training) & $1, 2, 4$ \\
                                                                &  Inner Loop Learning Rate & $0.05, 0.01, 0.1, 0.001$ \\ 
                                                                &  Meta-Path Inner Loop Steps (Training) & $1, 2, 4$ \\
                                                                &  Meta-Path Inner Loop Learning Rate & $0.05, 0.01, 0.1, 0.001$ \\ 
                                                                &  Outer Loop Learning Rate & $0.01, 0.005, 0.001, 0.0005$ \\
                                                                & Epochs & $30, 40, 50, 60$ \\
                                                                & Batch Size & $10, 20$ \\
        \midrule
		\multirow{9}{*}{MetaEmb \cite{10.1145/3331184.3331268}}&  Hidden Layers Size & $32, 64, 128$ \\   
                                                                &  Embedding Dimension & $32, 64, 128$ \\  
                                                                &  Inner Loop Learning Rate & $0.05, 0.01, 0.1, 0.001$ \\  
                                                                &  Learning Rate & $0.1, 0.05, 0.001$ \\
                                                                & Epochs & $30, 40, 50, 60$ \\
                                                                & Batch Size & $10, 20$ \\
                                                                & Pre-Train Epochs & $40, 50$ \\
                                                                & Pre-Train Learning Rate & $0.001, 0.0005, 0.005, 0.01$ \\
                                                                & $\alpha$ & $0.1, 0.2$ \\
        \midrule
		\multirow{10}{*}{MWUF \cite{10.1145/3404835.3462843}}    &  Hidden Layers Size & $32, 64, 128$ \\   
                                                                &  Embedding Dimension & $32, 64, 128$ \\ 
                                                                &  Scale Network Hidden Dimension & $32, 64, 128$ \\ 
                                                                &  Shift Network Hidden Dimension & $32, 64, 128$ \\ 
                                                                &  Cold Loss Learning Rate & $0.05, 0.01, 0.005$ \\
                                                                &  Learning Rate & $0.01, 0.05, 0.1, 0.0005$ \\
                                                                & Epochs & $30, 40, 50, 60$ \\
                                                                & Batch Size & $10, 20$ \\
                                                                & Pre-Train Epochs & $40, 50$ \\
                                                                & Pre-Train Learning Rate & $0.001, 0.005, 0.01$ \\
        \midrule
		\multirow{11}{*}{MAMO \cite{10.1145/3394486.3403113}}    &  Hidden Layers Size & $32, 64, 128$ \\   
                                                                &  Embedding Dimension & $32, 64, 128$ \\   
                                                                &  User Preference Types  & $1, 2, 3$ \\
                                                                &  Learning Rate & $0.01, 0.005, 0.001, 0.0005$ \\
                                                                & Epochs & $30, 40, 50, 60$ \\
                                                                & Batch Size & $10, 20$ \\
                                                                & $\alpha$ & $0.4, 0.5, 0.6$ \\
                                                                & $\beta$ & $0.1, 0.2$ \\
                                                                &  $\gamma$ & $0.1, 0.2$ \\
                                                                &  $\tau$ & $0.1, 0.2$ \\
                                                                &  $\rho$ & $0.001, 0.005, 0.01$ \\
        \midrule
        \multirow{9}{*}{TSCS \cite{Neupane_Zheng_Kong_Yu_2022}}    &  Hidden Layers Size & $32, 64, 128$ \\   
                                                                &  Embedding Dimension & $32, 64, 128$ \\   
                                                                &  User Embedding Dimension & $32, 64, 128$ \\
                                                                &  Inner Loop Steps (Training) & $1, 2, 4$ \\
                                                                &  Inner Loop Steps (Testing) & $1, 3, 5, 8$ \\
                                                                &  Inner Loop Learning Rate & $0.05, 0.01, 0.1, 0.001$ \\ 
                                                                & Epochs & $30, 40, 50, 60$ \\
                                                                & Batch Size & $10, 20$ \\
                                                                & Outer Loop Learning Rate & $0.01, 0.001, 0.0001, 0.0005$ \\
		\bottomrule
	\end{tabular}
	}
\end{table}

\section{Hyperparameter Tuning}\label{appendix:baselines_hyperparam}
\begin{table}
	\caption{Values used for the hyperparamter tuning of the baseline non-meta-learning models.} 
	\label{tab:baseline_non_meta_hyp}
	\centering
	\resizebox{\columnwidth}{!}{%
	\begin{tabular}{lcc}
		\toprule
		Model                                  & Hyperparameter & Values \\
		\midrule
		\multirow{7}{*}{DeepFM \cite{Guo2017DeepFMAF}}    &  Hidden Layers Size & $32, 64, 128$ \\   
                                                          &  Embedding Dimension & $32, 64, 128$ \\   
                                                          &  User Embedding Dimension & $32, 64, 128$ \\
                                                          &  Learning Rate & $0.01, 0.005, 0.001, 0.0005$ \\
                                                                & Dropout Probability & $0.0, 0.2, 0.1$ \\
                                                                & Epochs & $30, 40, 50, 60$ \\
                                                                & Batch Size & $10, 20$ \\
        \midrule
		\multirow{10}{*}{DropoutNet \cite{NIPS2017_dbd22ba3}}    &  Hidden Layers Size & $32, 64, 128$ \\   
                                                                &  Embedding Dimension & $32, 64, 128$ \\   
                                                                &  User Embedding Dimension & $32, 64, 128$ \\
                                                                &  Learning Rate & $0.01, 0.005, 0.001, 0.0005$ \\
                                                                & Dropout Probability & $0.0, 0.2, 0.1$ \\
                                                                & Epochs & $30, 40, 50, 60$ \\
                                                                & Batch Size & $10, 20$ \\
                                                                & Content Dropout Probability & $0.1, 0.2, 0.3, 0.4$ \\
                                                                & Matrix Factorization Model Learning Rate & $0.001, 0.005$ \\
                                                                & Matrix Factorization Model Training Epochs & $40, 50, 60$ \\
		\bottomrule
	\end{tabular}
	}
\end{table}

\begin{table}
	\caption{Values used for the hyperparamter tuning of our proposed model.} 
	\label{tab:ours_hyp}
	\centering
	\resizebox{\columnwidth}{!}{%
	\begin{tabular}{lcc}
		\toprule
		Model                                  & Hyperparameter & Values \\
		\midrule
		\multirow{13}{*}{Ours}    & Hidden Layers Size & $32, 64, 128$ \\  
                                                          & User Representation Dimension & $16, 32, 64, 128$ \\
                                                          & Item Representation Dimension & $16, 32, 64, 128$ \\
                                                          & Learning Rate & $0.1, 0.01, 0.005, 0.001$ \\
                                                          & L2 Regularization & $0.0, 0.1, 0.01, 0.001$ \\
                                                          & Dropout Probability & $0.0, 0.2, 0.1, 0.3$ \\
                                                          & Epochs & $5, 10, 20, 30, 40, 50, 60$ \\
                                                          & Batch Size & $10, 20$ \\
                                                          & Number of GNN layers & $1, 2, 3$ \\
                                                          & Batch Normalization & True, False \\
                                                          & Social Connections & True, False \\
                                                          & ``Same-rating'' graph & True, False \\
                                                          & ``Same-attribute'' graph & True, False \\
		\bottomrule
	\end{tabular}
	}
\end{table}

To tune the hyperparameters for the considered models, we perform 100 steps of random search over the values presented in Table \ref{tab:baseline_non_meta_hyp} for the baseline non-meta-learning methods, in Table \ref{tab:baseline_hyp} for the meta-learning models, and in Table \ref{tab:ours_hyp} for our model. For each configuration we train the model, and test it on the validation set. We then take the configuration with the lowest loss on the validation set. 

To decide which values to explore, we base ourselves on the hyperparameters values provided by the authors in their paper, and/or on the values from the Recbole repository \cite{recbole,recbole[1.1.1],recbole[2.0]}.

\end{document}